\documentclass[
 aip,
 amsmath,amssymb,
 twocolumn,
]{revtex4}

\usepackage{textcase}
\usepackage{graphicx}
\usepackage{dcolumn}
\usepackage{bm}
\usepackage{amsmath,amsthm,amssymb,mathrsfs}

\begin{document}

\title{Maximizing Spin Torque Diode Voltage by Optimizing Magnetization Alignment}

\author{Tomohiro Taniguchi}
\author{Hiroshi Imamura}

\affiliation{ 
Spintronics Research Center, AIST, 1-1-1 Umezono, Tsukuba 305-8568, Japan
}

\date{\today}%

\begin{abstract}
      The optimum condition of the magnetization alignment to maximize the spin torque diode voltage is derived 
      by solving the Landau-Lifshitz-Gilbert equation. 
      We show that the optimized diode voltage can be one order of magnitude larger than 
      that of the conventional alignment where the easy axes of the free and the pinned layers are parallel. 
      These analytical predictions are confirmed by numerical simulations. 
\end{abstract}

\maketitle


There has been great interest in spin-torque-induced magnetization dynamics \cite{slonczewski89,slonczewski96,slonczewski02,berger96} 
due to its potential application to spintronics devices 
such as magnetic random access memory (MRAM) and microwave oscillators. 
A spin torque diode \cite{tulapurkar05,kubota08,sankey08,suzuki08,yakata09,wang09,ishibashi10,miwa12,bang12} 
is another important spintronics application, 
which enables us to rectify an alternating current in magnetic tunnel junction (MTJ) 
by synchronizing the current with the resonant oscillation of 
the tunnel magnetoresistance (TMR). 
In 2010, a spin torque diode effect with relatively large sensitivity ($\sim 170$ mV/mW) 
was observed experimentally \cite{ishibashi10}; 
however, the observed sensitivity was still lower than that of the Schottky diode. 


The spin torque diode effect arises from the combination of the spin torque and the TMR effects. 
The spin torque originating from an alternating current 
induces a small oscillation of the magnetization of the free layer around its steady state, 
as a result of which the resistance of the MTJ oscillates through the TMR effect. 
The oscillations of the current and the resistance create a direct voltage 
called a spin torque diode voltage. 
A large diode voltage, 
which determines the sensitivity of the diode, is obtained 
at the resonance frequency of the free layer. 
Hereafter, we refer to the MTJ in which 
the magnetization of the pinned layer is parallel to the easy axis of the free layer 
as the conventional alignment \cite{tulapurkar05,kubota08}. 
The diode voltage in the conventional alignment is on the order of 10 - 100 $\mu$V. 
A further increase of the diode voltage is desirable 
to excess the sensitivity of the Schottky diode. 


In this letter, 
we show that the spin torque diode voltage can be significantly enhanced 
by choosing an appropriate magnetization alignment of the free and pinned layers. 
We derive the optimum condition to maximize the diode voltage 
by solving the Landau-Lifshitz-Gilbert (LLG) equation. 
The optimum condition is determined by 
the competition between 
the contributions from the amplitude of the TMR oscillation 
and the linewidth of the power spectrum of the magnetization oscillation. 
We show that the optimum alignment shifts from the orthogonal alignment, 
and that the diode voltage with the optimized condition 
can be one order of magnitude larger than 
that of the conventional alignment. 
These analytical predictions are confirmed by numerically solving the LLG equation. 



\begin{figure}
\centerline{\includegraphics[width=0.6\columnwidth]{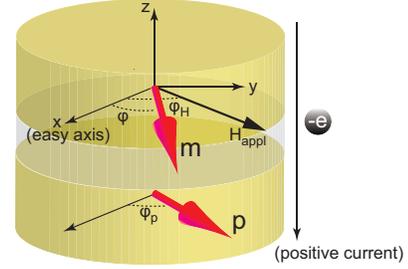}}
\caption{
         A schematic view of the spin torque diode system. 
         The positive current is defined as the electron flow 
         from the free to the pinned layer. 
         The easy axis of the free layer is parallel to the $x$-axis. 
         The unit vectors pointing in the direction of the magnetizations 
         of the free and the pinned layers are denoted as 
         $\mathbf{m}$ and $\mathbf{p}$, respectively. 
         The angles $\varphi$, $\varphi_{H}$, and $\varphi_{\rm p}$ are 
         the steady state of the magnetization of the free layer 
         and the directions of the applied field and the magnetization of the pinned layer, respectively. 
         \vspace{-3ex}}
\label{fig:fig1}
\end{figure}



The system we consider is schematically shown in Fig. \ref{fig:fig1}. 
The MTJ consists of the free and the pinned layers separated by a nonmagnetic barrier. 
The $x$-axis is parallel to the easy axis of the free layer 
while the $z$-axis is normal to the film plane. 
The unit vectors pointing in the direction of the magnetizations 
of the free and the pinned layers are denoted as 
$\mathbf{m}$ and $\mathbf{p}=(\cos\varphi_{\rm p},\sin\varphi_{\rm p},0)$, respectively, 
where the direction of the magnetization of the pinned layer 
is described by the angle $\varphi_{\rm p}$. 
In the conventional alignment, 
$\varphi_{\rm p}=0$. 


The magnetization dynamics of the free layer is described by using 
the LLG equation with the spin torque 
\cite{slonczewski89,slonczewski96,slonczewski02,berger96,tulapurkar05,landau35,lifshitz80,gilbert04}, 
\begin{equation}
  \frac{{\rm d}\mathbf{m}}{{\rm d}t}
  =
  -\gamma
  \mathbf{m}
  \times
  \mathbf{H}
  -
  \gamma
  a_{J}
  \mathbf{m}
  \times
  \left(
    \mathbf{p}
    \times
    \mathbf{m}
  \right)
  +
  \gamma
  b_{J}
  \mathbf{p}
  \times
  \mathbf{m}
  +
  \alpha
  \mathbf{m}
  \times
  \frac{{\rm d}\mathbf{m}}{{\rm d}t}.
  \label{eq:LLG}
\end{equation}
Throughout this letter, 
we assume that the magnetization dynamics is well-described by the macrospin model. 
The magnetic field $\mathbf{H}=(H_{\rm appl}\cos\varphi_{H}+H_{\rm K}m_{x},H_{\rm appl}\sin\varphi_{H},-4\pi M m_{z})$ 
is defined by the derivative the magnetic energy density, 
$E=-MH_{\rm appl}(\cos\varphi_{H}m_{x}+\sin\varphi_{H}m_{y})-(MH_{\rm K}/2)m_{x}^{2}+2\pi M^{2}m_{z}^{2}$, 
with respect to $\mathbf{m}$, 
and consists of the applied field, $H_{\rm appl}$, 
the uniaxial anisotropy field along the easy axis, $H_{\rm K}$, 
and the demagnetization field along the hard axis, $4\pi M$. 
The gyromagnetic ratio and Gilbert damping constant are denoted as 
$\gamma$ and $\alpha$, respectively. 
The spin torque \cite{tulapurkar05,kubota08} consists of the Slonczewski torque, $a_{J}$, 
and the field like torque, $b_{J}$, defined as 
\begin{equation}
  a_{J}
  =
  \frac{\hbar \eta I}{2eMV}, 
\end{equation}
and $b_{J}=\beta a_{J}$, respectively, 
where $\beta$ is the ratio between the Slonczewski torque 
and the field like torque. 
Here, $M$ and $V$ are the saturation magnetization and the volume of the free layer, respectively. 
The spin polarization of the current is denoted as $\eta$. 
The current $I$ consists of the direct and the alternating currents, 
due to which $a_{J}$ and $b_{J}$ are decomposed into dc and ac parts as
$a_{J}=a_{J({\rm dc})}+a_{J({\rm ac})}$ and $b_{J}=b_{J({\rm dc})}+b_{J({\rm ac})}$, respectively. 
The positive current is defined as the electron flow 
from the free to the pinned layer. 


The spin torque diode effect arises from 
the small amplitude oscillation of the magnetization of the free layer 
around the steady state. 
Let us introduce two angles, $(\theta,\varphi)$, 
characterizing the direction of the magnetization at the steady state 
as $\mathbf{m}^{(0)}=(\sin\theta\cos\varphi,\sin\theta\sin\varphi,\cos\theta)$. 
From eq. (\ref{eq:LLG}),
the steady state should satisfy the following two conditions \cite{vonsovskii66}: 
\begin{equation}
\begin{split}
  &
  H_{\rm appl}
  \cos\theta
  \cos(\varphi-\varphi_{H})
  +
  \left(
    H_{\rm K}
    \cos^{2}\varphi
    +
    4\pi M
  \right)
  \sin\theta
  \cos\theta
\\
  &-
  a_{J({\rm dc})}
  \sin(\varphi-\varphi_{\rm p})
  +
  b_{J({\rm dc})}
  \cos\theta
  \cos(\varphi-\varphi_{\rm p})
  =
  0,
  \label{eq:eq_condition_1}
\end{split}
\end{equation}
\begin{equation}
\begin{split}
&
  H_{\rm appl}
  \sin(\varphi-\varphi_{H})
  +
  H_{\rm K}
  \sin\theta
  \sin\varphi
  \cos\varphi
\\
  &-
  a_{J({\rm dc})}
  \cos\theta
  \cos(\varphi-\varphi_{\rm p})
  -
  b_{J({\rm dc})}
  \sin(\varphi-\varphi_{\rm p})
  =
  0.
  \label{eq:eq_condition_2}
\end{split}
\end{equation}
In the absence of a direct current, 
the steady state is equal to the equilibrium state, 
i.e., the minimum state of the magnetic energy density $E$ 
located in the film plane ($\theta=\pi/2$). 
Below, we set $\theta=\pi/2$ 
by assuming that the magnitudes of $a_{J({\rm dc})}$ and $b_{J({\rm dc})}$ are small 
compared with $|\mathbf{H}|$ \cite{comment1}. 
Let us introduce the coordinates $(X,Y,Z)$ 
in which the $Z$-axis is parallel to the steady state $\mathbf{m}^{(0)}$. 
The LLG equation can be linearized 
by applying the approximations $m_{Z} \simeq 1$ and $|m_{X}|,|m_{Y}| \ll 1$. 
The linearized LLG equation is given by 
\begin{equation}
  \frac{1}{\gamma}
  \frac{{\rm d}}{{\rm d}t}
  \begin{pmatrix}
    m_{X} \\
    m_{Y} 
  \end{pmatrix}
  +
  \mathsf{M}
  \begin{pmatrix}
    m_{X} \\
    m_{Y}
  \end{pmatrix}
  =
  \begin{pmatrix}
    b_{J({\rm ac})} p_{Y} \\
    -a_{J({\rm ac})} p_{Y} 
  \end{pmatrix}, 
  \label{eq:LLG_linear}
\end{equation}
where we use the approximation $1+\alpha^{2} \simeq 1$ 
because the Gilbert damping constant $\alpha$ is on the order of $10^{-2}$ \cite{oogane06}. 
The components of the coefficient matrix $\mathsf{M}$ are given by 
\begin{equation}
  \mathsf{M}_{11}
  =
  -a_{J({\rm dc})}
  p_{Z}
  +
  \alpha
  \left(
    H_{X}
    +
    b_{J({\rm dc})}
    p_{Z}
  \right),
\end{equation}
\begin{equation}
  \mathsf{M}_{12}
  =
  \left(
    H_{Y}
    +
    b_{J({\rm dc})}
    p_{Z}
  \right)
  +
  \alpha
  a_{J}({\rm dc})
  p_{Z},
\end{equation}
\begin{equation}
  \mathsf{M}_{21}
  =
  -\left(
    H_{X}
    +
    b_{J({\rm dc})}
    p_{Z}
  \right)
  -
  \alpha 
  a_{J({\rm dc})}
  p_{Z},
\end{equation}
\begin{equation}
  \mathsf{M}_{22}
  =
  -a_{J({\rm dc})}
  p_{Z}
  +
  \alpha
  \left(
    H_{Y}
    +
    b_{J({\rm dc})}
    p_{Z}
  \right),
\end{equation}
where $H_{X}$ and $H_{Y}$ are given by 
\begin{equation}
  H_{X}
  =
  H_{\rm appl}
  \cos(\varphi-\varphi_{H})
  +
  H_{\rm K}
  \cos^{2}\varphi
  +
  4\pi M,
\end{equation}
\begin{equation}
  H_{Y}
  =
  H_{\rm appl}
  \cos(\varphi-\varphi_{H})
  +
  H_{\rm K}
  \cos 2\varphi.
\end{equation} 
The components of $\mathbf{p}$ in the $XYZ$-coordinate are given by 
$(p_{X},p_{Y},p_{Z})=(0,-\sin(\varphi-\varphi_{\rm p}),\cos(\varphi-\varphi_{\rm p}))$. 
We assume that the alternating current is given by 
$I_{\rm ac}\sin(2\pi ft)$. 
Then, the solutions of $m_{X}$ and $m_{Y}$ can be obtained by solving eq. (\ref{eq:LLG_linear}). 
The explicit forms of $m_{X}$ and $m_{Y}$, respectively, are given by 
\begin{equation}
\begin{split}
  m_{X}
  =&
  -{\rm Im}
  \left[
    \frac{\tilde{\gamma}^{2} ( H_{Y} + b_{J({\rm dc})}p_{Z}) p_{Y}}{f^{2}-f_{\rm res}^{2}-{\rm i}f \Delta f}
    {\rm e}^{2\pi {\rm i}ft}
  \right]
  \tilde{a}_{J({\rm ac})}
\\
  &-
  {\rm Im}
  \left[
    \frac{\tilde{\gamma} ( {\rm i}f - \tilde{\gamma} a_{J({\rm dc})}p_{Z}) p_{Y}}{f^{2}-f_{\rm res}^{2}-{\rm i}f \Delta f}
    {\rm e}^{2\pi {\rm i}ft}
  \right]
  \tilde{b}_{J({\rm ac})},
  \label{eq:mX}
\end{split}
\end{equation}
\begin{equation}
\begin{split}
  m_{Y}
  =&
  {\rm Im}
  \left[
    \frac{\tilde{\gamma} ( {\rm i}f - \tilde{\gamma} a_{J({\rm dc})}p_{Z}) p_{Y}}{f^{2}-f_{\rm res}^{2}-{\rm i}f \Delta f}
    {\rm e}^{2\pi {\rm i}ft}
  \right]
  \tilde{a}_{J({\rm ac})},
\\
  &-
  {\rm Im}
  \left[
    \frac{\tilde{\gamma}^{2} ( H_{X} + b_{J({\rm dc})}p_{Z}) p_{Y}}{f^{2}-f_{\rm res}^{2}-{\rm i}f \Delta f}
    {\rm e}^{2\pi {\rm i}ft}
  \right]
  \tilde{b}_{J({\rm ac})},
  \label{eq:mY}
\end{split}
\end{equation}
where $\tilde{\gamma}=\gamma/(2\pi)$, and 
$\tilde{a}_{J({\rm ac})}$ and $\tilde{b}_{J({\rm ac})}$ are defined as 
$a_{J({\rm ac})}=\tilde{a}_{J({\rm ac})}\sin(2\pi ft)$ 
and $b_{J({\rm ac})}=\tilde{b}_{J({\rm ac})}\sin(2\pi ft)$, respectively. 
The resonant frequency $f_{\rm res}$ and 
the linewidth $\Delta f$ \cite{comment2} are defined as 
\begin{equation}
  f_{\rm res}
  =
  \frac{\gamma}{2\pi}
  \sqrt{
    \left(
      H_{X}
      +
      b_{J({\rm dc})}
      p_{Z}
    \right)
    \left(
      H_{Y}
      +
      b_{J({\rm dc})}
      p_{Z}
    \right)
    +
    (a_{J({\rm dc})}p_{Z})^{2}
  },
  \label{eq:res}
\end{equation}
\begin{equation}
  \Delta f 
  =
  \frac{\gamma}{2\pi}
  \left[
    \alpha
    \left(
      H_{X}
      +
      H_{Y}
    \right)
    -
    2 a_{J({\rm dc})}
    p_{Z}
  \right].
  \label{eq:linewidth}
\end{equation}
In the absence of a direct current, 
$f_{\rm res}$ is identical to the ferromagnetic resonant (FMR) frequency, $f_{\rm FMR}$. 

The spin torque diode voltage is defined as 
\begin{equation}
\begin{split}
  V_{\rm dc}
  &=
  \frac{1}{T}
  \int_{0}^{T} {\rm d}t 
  R I_{\rm ac} \sin(2\pi ft)
\\
  &=
  \frac{1}{T}
  \int_{0}^{T} {\rm d}t
  \left[
    R_{\rm P}
    +
    \frac{\Delta R}{2}
    \left(
      1
      -
      \mathbf{m}
      \cdot
      \mathbf{p}
    \right)
  \right]
  I_{\rm ac}
  \sin(2\pi ft),
  \label{eq:voltage_def}
\end{split}
\end{equation}
where $T=1/f$ 
and $R$ is the TMR given by $R=R_{\rm P}+(\Delta R/2)(1-\mathbf{m}\cdot\mathbf{p})$,
where $\Delta R = R_{\rm AP}-R_{\rm P}$ is the difference between the resistances 
at the parallel ($R_{\rm P}$) and antiparallel ($R_{\rm AP}$) alignments of the magnetizations. 
It should be noted that 
only the oscillation part of the TMR, 
$-(\Delta R/2) \mathbf{m}\cdot\mathbf{p}$, 
where $m_{X}$ and $m_{Y}$ oscillate with the frequency $f$, 
contributes to the diode voltage. 
By substituting eqs. (\ref{eq:mX}) and (\ref{eq:mY}) into eq. (\ref{eq:voltage_def}),
the explicit form of $V_{\rm dc}$ is given by
\begin{equation}
  V_{\rm dc}
  =
  \frac{\Delta R I_{\rm ac}}{4}
  \left[
    \mathscr{L}(f)
    +
    \mathscr{A}(f)
  \right].
  \label{eq:voltage}
\end{equation}
Here, the Lorentzian and anti-Lorentzian parts, $\mathscr{L}(f)$ and $\mathscr{A}(f)$, are given by 
\begin{equation}
  \mathscr{L}(f)
  =
  \frac{f^{2} \Delta f \tilde{\gamma} \tilde{a}_{J({\rm ac})}(1-p_{Z}^{2})}
    {(f^{2}-f_{\rm res}^{2})^{2} + (f \Delta f)^{2}}, 
\end{equation}
\begin{equation}
  \mathscr{A}(f)
  =
  -\frac{(f^{2}-f_{\rm res}^{2})\tilde{\gamma}^{2} (\mathscr{H}_{a} \tilde{a}_{J({\rm ac})} - \mathscr{H}_{b} \tilde{b}_{J({\rm ac})})}
    {(f^{2}-f_{\rm res}^{2})^{2} + (f \Delta f)^{2}}, 
\end{equation}
where $\mathscr{H}_{a}=-a_{J({\rm dc})}p_{Z}p_{Y}^{2}$ and $\mathscr{H}_{b}=(H_{X} + b_{J({\rm dc})}p_{Z})p_{Y}^{2}$. 



\begin{figure}
\centerline{\includegraphics[width=1.0\columnwidth]{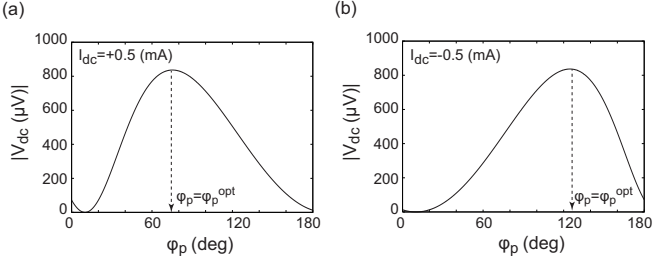}}
\caption{
         Dependence of the magnitude of the diode voltage, eq. (\ref{eq:voltage_tmp}), 
         on the direction of the magnetization of the pinned layer. 
         The values of the direct current are (a) $I_{\rm dc}=0.5$ mA and (b) $I_{\rm dc}=-0.5$ mA. 
         The conventional alignment corresponds to $\varphi_{\rm p}=0$. 
         The voltage is maximized at the optimized angle $\varphi_{\rm p}^{\rm opt}$, 
         and is zero at $\varphi_{\rm p}=\varphi$. 
         \vspace{-3ex}}
\label{fig:fig2}
\end{figure}




The diode voltage, eq. (\ref{eq:voltage}), 
shows a peak near the resonant frequency $f_{\rm res}$. 
At $f=f_{\rm res}$, 
$V_{\rm dc}$ is given by \cite{suzuki08} 
\begin{equation}
  V_{\rm dc}(f_{\rm res})
  =
  \frac{\Delta R I_{\rm ac}}{4}
  \frac{\tilde{a}_{J({\rm ac})}\sin^{2}(\varphi-\varphi_{\rm p})}{\alpha(H_{X} + H_{Y}) - 2 a_{J({\rm dc})}\cos(\varphi-\varphi_{\rm p})},
  \label{eq:voltage_tmp}
\end{equation}
where we use $p_{Z}=\cos(\varphi-\varphi_{\rm P})$. 
The term $\sin^{2}(\varphi-\varphi_{\rm p})$ in the numerator of eq. (\ref{eq:voltage_tmp}) arises 
from 
the oscillation part of the TMR, 
$-(\Delta R/2)\mathbf{m}\cdot\mathbf{p}$, 
and is maximized in the orthogonal alignment of the magnetizations, 
$\varphi_{\rm p}-\varphi=90^{\circ}$. 
The maximum diode voltage has been estimated in this orthogonal alignment \cite{suzuki08}. 
However, the diode voltage depends on not only the TMR 
but also the linewidth of the power spectrum of $m_{X}$ and $m_{Y}$. 
The term $2a_{J({\rm dc})}\cos(\varphi-\varphi_{\rm p})$ in the denominator of eq. (\ref{eq:voltage_tmp}) represents 
the enhancement or the reduction of the linewidth 
due to the spin torque acting as a damping or anti-damping factor, depending on the direction of the current. 
The optimum condition is determined by the competition between 
the contributions from the amplitude of the TMR oscillation and 
the linewidth of the power spectrum of the magnetization oscillation. 
We find that the diode voltage, $V_{\rm dc}(f_{\rm res})$, can be maximized 
when the magnetization of the pinned layer points to the direction 
\begin{equation}
  \varphi_{\rm p}^{\rm opt}
  =
  \cos^{-1}
  \left[
    \frac{I_{\rm c}}{I_{\rm dc}}
    \mp
    \sqrt{
      \left(
        \frac{I_{\rm c}}{I_{\rm dc}}
      \right)^{2}
      -
      1
    }
  \right]
  +
  \varphi,
  \label{eq:varphi_p_opt}
\end{equation}
where the double sign "$\mp$" means 
the upper ($-$) for $I_{\rm dc}/I_{\rm c}>0$ and the lower ($+$) for $I_{\rm dc}/I_{\rm c}<0$. 
The quantity $I_{\rm c}$ is the absolute value of the critical current 
of the spin-torque-induced magnetization dynamics around the steady state defined as 
\begin{equation}
  I_{\rm c}
  =
  \frac{2 \alpha eMV}{\hbar\eta}
  \left[
    H_{\rm appl}
    \cos(\varphi-\varphi_{H})
    +
    H_{\rm K}
    \frac{\cos^{2}\varphi + \cos 2\varphi}{2}
    +
    2 \pi M
  \right].
  \label{eq:Ic}
\end{equation}
The optimum alignment, $\varphi_{\rm p}^{\rm opt}-\varphi$, shifts from the orthogonal alignment 
as long as the direct current is finite, 
while $\varphi_{\rm p}^{\rm opt}-\varphi=90^{\circ}$ for $I_{\rm dc}=0$ 
because the spin torque does not affect the linewidth in this case. 
The maximized diode voltage, $V_{\rm dc}^{\rm opt}(f_{\rm res})$, is given by 
\begin{equation}
  V_{\rm dc}^{\rm opt}(f_{\rm res})
  =
  \frac{\Delta R I_{\rm ac}^{2}}{4 I_{\rm dc}}
  \left[
    \frac{I_{\rm c}}{I_{\rm dc}}
    \mp
    \sqrt{
      \left(
        \frac{I_{\rm c}}{I_{\rm dc}}
      \right)^{2}
      -
      1
    }
  \right],
  \label{eq:voltage_max}
\end{equation}
where the meaning of the double sign $"\mp"$ is the same as in eq. (\ref{eq:varphi_p_opt}). 
Equations (\ref{eq:varphi_p_opt}) and (\ref{eq:voltage_max}) are the main results of this study. 
These results indicate that the spin torque diode voltage can be significantly enhanced 
by choosing an appropriate alignment of the magnetizations. 
It should be noted that 
since $\varphi_{\rm p}^{\rm opt}$ is a real number, 
$|I_{\rm dc}|$ should be less than $|I_{\rm c}|$. 
In the opposite case, $I_{\rm dc}/I_{\rm c} \ge 1$, 
the above formula is not applicable 
because the spin torque equals or overcomes the damping, 
due to which the steady state becomes unstable, 
and thus, the LLG equation cannot be linearized. 
In the limit of $I_{\rm dc}\to 0$, 
eq. (\ref{eq:voltage_max}) becomes 
$\Delta R I_{\rm ac}^{2}/(8 I_{\rm c})$. 




\begin{figure}
\centerline{
\includegraphics[width=1.0\columnwidth]{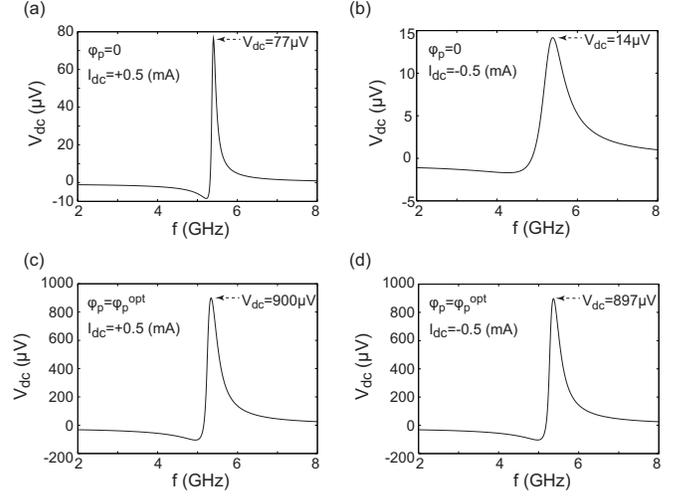}}
\caption{
         Spin torque diode voltages obtained by numerically solving the LLG equation, 
         where the direction of the magnetization of the pinned layers 
         and the value of the direct current are taken to be 
         (a) $(\varphi_{\rm p},I_{\rm dc}({\rm mA}))=(0,0.5)$, 
         (b) $(0,-0.5)$, 
         (c) $(\varphi_{\rm p}^{\rm opt},0.5)$, 
         and (d) $(\varphi_{\rm p}^{\rm opt},-0.5)$, respectively. 
         \vspace{-3ex}}
\label{fig:fig3}
\end{figure}



Let us quantitatively estimate how much the spin torque diode voltage can be enhanced 
by the optimization of the magnetization alignment. 
Figures \ref{fig:fig2}(a) and \ref{fig:fig2}(b) show the dependence of 
the magnitudes of $V_{\rm dc}(f_{\rm res})$ 
on the direction of the magnetization of the pinned layer, $\varphi_{\rm p}$, 
where $I_{\rm dc}=0.5$ mA in (a) and $-0.5$ mA in (b). 
The values of the other parameters are taken to be 
$M=1000$ emu/c.c., $H_{\rm K}=200$ Oe, $H_{\rm appl}=100$ Oe, $\varphi_{H}=30^{\circ}$, $V=\pi \times 80 \times 35 \times 2$ nm${}^{3}$, 
$\gamma=17.64$ MHz/Oe, $\alpha=0.01$, $\eta=0.5$, $\beta=0.1$, 
$I_{\rm ac}=0.2$ mA, and $\Delta R = 100$ $\Omega$. 
The equilibrium direction of the free layer is estimated to be $\varphi \simeq 10^{\circ}$ 
while the optimized directions of the pinned layer are 
$\varphi_{\rm p}^{\rm opt} \simeq 75^{\circ}$ for $I_{\rm dc}=0.5$ mA 
and $125^{\circ}$ for $I_{\rm dc}=-0.5$ mA. 
It should be noted that the optimized alignment, $\varphi_{\rm p}^{\rm opt}-\varphi$, 
shifts from the orthogonal alignment. 
In the conventional alignment ($\varphi_{\rm p}=0$), 
the diode voltages are 72 $\mu$V for $I_{\rm dc}=0.5$ mA and 13 $\mu$V for $I_{\rm dc}=-0.5$ mA, 
where the diode voltage for $I_{\rm dc}>0$ is larger than that for $I_{\rm dc}<0$ 
because the positive current has an anti-damping effect in this case, 
and thus, reduces the linewidth. 
On the other hand, the magnitude of the maximum diode voltage $V_{\rm dc}^{\rm opt}$ is 
estimated to be 837 $\mu$V for $I_{\rm dc}=\pm 0.5$ mA. 
Thus, the diode voltage satisfying the optimum condition is expected to be 
one order of magnitude larger than 
that of the conventional alignment. 


We confirmed the above analytical predictions 
by numerically solving the LLG equation \cite{taniguchi11}. 
Figures \ref{fig:fig3}(a) and \ref{fig:fig3}(b) show the dependences of the diode voltages 
on the frequency of the alternating current 
at $\varphi_{\rm p}=0$ with (a) $I_{\rm dc}=0.5$ and (b) $-0.5$ mA, respectively, 
while Figs. \ref{fig:fig3}(c) and \ref{fig:fig3}(d) show 
$V_{\rm dc}$ at $\varphi_{\rm p}=\varphi_{\rm p}^{\rm opt}$ with (c) $I_{\rm dc}=0.5$ and (d) $-0.5$ mA.
Although the direct current affects the resonant frequency, as shown in eq. (\ref{eq:res}), 
the peak of the diode voltage appears approximately at the FMR frequency, $f_{\rm FMR}=5.3$ GHz, 
because the magnitude of the direct current is relatively small; 
the peak frequencies of Fig. \ref{fig:fig3}(a) and \ref{fig:fig3}(b) are 5.4 GHz 
while those of Fig. \ref{fig:fig3}(c) and \ref{fig:fig3}(d) are 5.3 GHz.  
The magnitudes of the maximum voltage in Figs. \ref{fig:fig3}(a) and \ref{fig:fig3}(b) are 
77 and 14 $\mu$V, 
while those of Figs. \ref{fig:fig3}(c) and \ref{fig:fig3}(d) are 900 and 897 $\mu$V, respectively. 
These results have a good agreement with Fig. \ref{fig:fig2}, 
showing the validity of eqs. (\ref{eq:varphi_p_opt}) and (\ref{eq:voltage_max}). 


In conclusion, 
we derived the optimum condition of the magnetization alignment of the free and the pinned layers 
to maximize the spin torque diode voltage 
by analyzing the competition between the oscillation of the tunneling magnetoresistance 
and the reduction of the linewidth due to the spin torque. 
We showed that the optimum alignment shifts from the orthogonal alignment. 
We also showed that, under the optimized condition, 
the diode voltage can be one order of magnitude larger that 
that in the conventional alignment. 
These analytical predictions were confirmed by numerical simulations. 
The results indicate that the diode voltage can be significantly enhanced 
by choosing an appropriate magnetization alignment. 
Experimentally, the direction of the magnetization of the pinned layer maybe controlled 
during the annealing process, 
as done in a TMR head \cite{comment3}.


The authors would like to acknowledge 
H. Kubota, H. Maehara, and S. Miwa 
for the valuable discussions they had with us. 




\begin{thebibliography}{22}
\bibitem[{com({\natexlab{a}})}]{comment1}
\bibinfo{note}{In the experiments \cite{kubota08}, the magnitude of the direct
  current is comparable to or smaller than that for the spin torque switching;
  see also eq. (\ref{eq:Ic}). It is well known that the critical current of the
  spin torque switching is on the order of $\alpha |\mathbf{H}|$. Since the
  Gilbert damping constant is very small $(\alpha \sim 10^{-2})$
  \cite{oogane06}, we assume that the effect of the direct current on the
  determination of $(\theta,\varphi)$ is negligible.}
\bibitem[{com({\natexlab{b}})}]{comment2}
\bibinfo{note}{We neglect the term $b_{J({\rm dc})}$ in eq.
  (\ref{eq:linewidth}) because the field like torque is smaller than the
  Slonczewki torque, and the magnitude of the Slonczewki torque is on the order
  of $\alpha |\mathbf{H}|$, as mentioned in ref. \cite{comment1}.}
\bibitem[{com({\natexlab{c}})}]{comment3}
\bibinfo{note}{The magnetization alignment of a TMR head of a hard disk drive
  is set to be orthogonal by annealing process.}
\end{thebibliography}



\end{document}